\documentclass{emulateapj}
\usepackage{color,amsmath}
\newcommand{\fAGN}{\ensuremath{f({\rm AGN})_{\rm MIR}}}
\newcommand{\fTOT}{\ensuremath{f({\rm AGN})_{\rm IR}}}

\begin{document}

\title{The AGN-Star Formation Connection: Future Prospects with JWST
}
\author{Allison Kirkpatrick\altaffilmark{1}, Stacey Alberts\altaffilmark{2}, Alexandra Pope\altaffilmark{3}, Guillermo Barro\altaffilmark{4}, Matteo Bonato\altaffilmark{5}, Dale D. Kocevski\altaffilmark{6}, Pablo P\'erez-Gonz\'alez\altaffilmark{7}, George H. Rieke\altaffilmark{2}, Lucia Rodr\'iguez-Mu\~noz\altaffilmark{8}, Anna Sajina\altaffilmark{9}, Norman Grogin\altaffilmark{10}, Kameswara Bharadwaj Mantha\altaffilmark{11}, 
Viraj Pandya\altaffilmark{12}, Janine Pforr\altaffilmark{13}, Paola Santini\altaffilmark{14}}
\altaffiltext{1}{Yale Center for Astronomy \& Astrophysics, Physics Department, P.O. Box 208120, New Haven, CT 06520, USA, allison.kirkpatrick@yale.edu}
\altaffiltext{2}{Steward Observatory, University of Arizona, 933 North Cherry Avenue, Tucson, AZ 85721, USA}
\altaffiltext{3}{Department of Astronomy, University of Massachusetts, Amherst, MA 01002, USA}
\altaffiltext{4}{ University of California, 501 Campbell Hall, Berkeley, CA 94720 Santa Cruz, USA}
\altaffiltext{5}{INAF-Osservatorio di Radioastronomia, Via Piero Gobetti 101, I-40129 Bologna, Italy}
\altaffiltext{6}{Department of Physics and Astronomy, Colby College, Waterville, ME 04901, USA}
\altaffiltext{7}{Departamento de Astrof\'isica y CC. de la Atm\'osfera, Universidad Complutense de Madrid, E-28040 Madrid, Spain}
\altaffiltext{8}{Dipartimento di Fisica e Astronomia, Universit\`a di Padova, vicolo dellOsservatorio 2, 35122 Padova, Italy}
\altaffiltext{9}{Department of Physics \& Astronomy, Tufts University, Medford, MA 02155, USA}
\altaffiltext{10}{Space Telescope Science Institute, 3700 San Martin Dr., Baltimore, MD 21218, USA}
\altaffiltext{11}{Department of Physics and Astronomy, University of Missouri-Kansas City, 5110 Rockhill Road, Kansas City, MO 64110, USA}
\altaffiltext{12}{UCO/Lick Observatory, Department of Astronomy and Astrophysics, University of California, Santa Cruz, CA 95064, USA}
\altaffiltext{13}{ESA/ESTEC, Keplerlaan 1, 2201 AZ Noordwijk, The Netherlands}
\altaffiltext{14}{INAF - Osservatorio Astronomico di Roma, via di Frascati 33, 00078 Monte Porzio Catone, Italy}

\begin{abstract}
The bulk of the stellar growth over cosmic time is dominated by IR luminous galaxies at cosmic noon ($z=1-2$), many of which harbor a hidden active galactic nucleus (AGN). 
We use state of the art infrared color diagnostics, combining {\it Spitzer} and {\it Herschel} observations, to separate dust-obscured AGN from dusty star
forming galaxies (SFGs) in the CANDELS and COSMOS surveys. We calculate 24 $\mu$m counts of SFGs, AGN/star forming ``Composites'', and AGN. AGN and Composites dominate the
counts above 0.8 mJy at 24 $\mu$m, and Composites form at least 25\% of an IR sample even to faint detection limits. We develop methods to use the Mid-Infrared Instrument (MIRI) on JWST to identify dust-obscured
AGN and Composite galaxies from $z\sim1-2$. With the sensitivity and spacing of MIRI filters, we will detect $>$4 times as many AGN hosts than with {\it Spitzer}/IRAC criteria. Any star formation rates based on the 7.7 $\mu$m PAH feature (likely to
be applied to MIRI photometry) must be corrected for the contribution of the AGN, or the SFR will be overestimated by
$\sim$35\% for cases where the AGN provides half the IR luminosity and $\sim$ 50\% when the AGN
accounts for 90\% of the luminosity.
Finally, we demonstrate that our MIRI color technique can select AGN with an Eddington ratio of $\lambda_{\rm Edd}\sim0.01$ and will identify AGN hosts with a higher
sSFR than X-ray techniques alone. JWST/MIRI will enable critical steps forward in identifying and understanding dust-obscured AGN and the link to their host galaxies. \end{abstract}

\section{Introduction}
The galaxies most actively contributing to the buildup of stellar mass at cosmic noon ($z\sim1-2$) contain large amounts of dust \citep[e.g.][and references therein]{murphy2011,madau2014}. This dust obscures the majority of star formation, making it necessary to study these galaxies through their dust emission at infrared wavelengths \citep{madau2014}. Additionally, the majority of supermassive black hole growth at these redshifts is also heavily dust-obscured \citep[e.g.,][]{hickox2007}. Many of the massive dusty galaxies contain a true mix of star formation and obscured black hole growth, the obscured signatures of which can be seen in their infrared spectral energy distribution (SED). These galaxies are then ideal laboratories for understanding the physical link between star formation and active galactic nuclei (AGN). The AGN-star formation connection is an open question, particularly whether AGN feedback is a key component of star formation quenching, and whether all galaxies have a distinct star formation phase followed by an AGN phase before ultimately quenching \citep[e.g.][]{sanders1988,hopkins2006}. The nature of AGN within strongly star forming galaxies (what we term ``Composites'') is even more uncertain. Do these objects represent a unique phase between star forming galaxies (SFGs) and AGN? Unfortunately, due to limitations of previous space telescopes, detailed studies of the energetics of these objects were severely restricted, but the {\it James Webb Space Telescope} (JWST) will reveal their true nature.

Prior to JWST, the most reliable method for identifying Composites and disentangling AGN emission from star formation was mid-IR spectroscopy from the {\it Spitzer Space Telescope}. The low resolution spectra can be modeled as a combination of star formation features (most notably the polycyclic aromatic hydrocarbons, or PAHs, that exist in photodissociation regions and in stellar/{\sc Hii} regions), and hot continuum emission primarily arising from a dusty torus surrounding the accreting black hole \citep{pope2008,coppin2010,kirkpatrick2012,sajina2012,caballero2015,kirkpatrick2015}. In this way, the division of IR luminosity between star formation and an AGN can be quantified. The medium-resolution spectrometer \citep[MRS;][]{wells2015}, which is part of the Mid-Infrared Instrument (MIRI) on JWST, will enable separation of PAH emission from continuum in the same manner, but with higher resolution and on smaller spatial scales within host galaxies. It will also enable detection of high ionization gas lines excited by the AGN \citep{bonato2017}, further improving our ability to detect and measure the physical properties (such as accretion rates and Eddington ratios) of dust-obscured black holes.

As there are only a few hundred {\it Spitzer} IRS spectra available for distant galaxies \citep{kirkpatrick2015}, color techniques were also developed to identify large samples of luminous dust-obscured AGN. 
The most popular color selection techniques are with {\it Spitzer} IRAC photometry \citep{lacy2004,stern2005,alonso2006,donley2012}, which separate AGN using different combinations of the 3.6, 4.5, 5.8, and 8.0\,$\mu$m filters. The original techniques presented in \citet{lacy2004} and \citet{stern2005} were limited to the most luminous AGN and become increasingly contaminated with galaxies when deeper IR data are used \citep{mendez2013}. Moreover, with increasing redshift, the rest wavelengths of these bands decrease, causing contamination of the AGN signatures by star forming galaxies to become significant such that the original IRAC-based criteria cannot be applied.
\citet{donley2012} propose more conservative IRAC criteria that, at cosmic noon, essentially separate galaxies that exhibit a so-called stellar bump (emission from stars that peaks at $\sim1.6\,\mu$m and then declines to a minimum around $\sim5\,\mu$m) from those that do not, where the torus radiation is strong enough to fill in the dip in the star forming spectrum around $3-5\,\mu$m, producing power-law emission such as is typical of unobscured AGN \citep[e.g.][]{elvis1994}. The \citet{donley2012} criteria increase the reliability of AGN color selection, although they are less complete due to excluding Composites, where the IR emission of the AGN does not dominate over the star formation.  

For the purposes of probing the AGN-star formation connection, the limitation of IRAC techniques is that AGN within strongly star forming galaxies can have different levels of host contamination. Then, many galaxies containing AGN signatures at longer wavelengths will also include a stellar bump and therefore be missed \citep{kirkpatrick2013,kirkpatrick2015}. To alleviate host contamination, \citet{messias2012} propose combining $K$-band with IRAC and 24\,$\mu$m to separate AGN from host galaxies all the way out to $z\sim7$. 
Going further, including mid-IR {\it and} far-IR colors can greatly improve the selection of Composite galaxies, since this will trace the contribution of warmer AGN-heated dust compared with cold dust from the diffuse interstellar medium in the host galaxy \citep{kirkpatrick2015}. However, this requires observations from the {\it Herschel Space Observatory}, which have a large beam size and do not reach the same depths as {\it Spitzer} observations. 
MIRI will greatly improve color selection techniques due to the increased sensitivity and the number of transmission filters covering the mid-infrared \citep{bouchet2015,glasse2015}. Now, we will be able to separate AGN from SFGs by comparing PAH emission with the minimum emission from stars that occurs around 5\,$\mu$m; in AGN, the stellar minimum is not visible due to strong torus emission, and Composites will lie in between strong AGN and pure SFGs in colorspace. 

In this paper, we build on the {\it Herschel} and {\it Spitzer} color selection techniques initially presented in \citet{kirkpatrick2013} to identify Composite galaxies at $z\sim1-2$ using the CANDELS and COSMOS surveys. We present galaxy counts of 24\,$\mu$m sources classified as SFGs, AGN, or Composites based on their IR colors, making this the first identified statistical sample of Composites at cosmic noon. We use this sample to predict black hole and star formation properties of samples that JWST/MIRI will identify. We also present color diagnostics for identifying both AGN and Composites using JWST/MIRI filters in three redshift bins. Throughout this paper, we assume a standard cosmology with  $H_{0}=70\,\rm{km}\,\rm{s}^{-1}\,\rm{Mpc}^{-1}$, $\Omega_{\rm{M}}=0.3$, and $\Omega_{\Lambda}=0.7$. 
 
\section{CANDELS and COSMOS Catalogs}
To calculate galaxy counts, we use {\it Spitzer} and {\it Herschel} photometry from the COSMOS, EGS, GOODS-S, and UDS fields from the Cosmic Assembly NearIR Deep Extragalactic Survey (CANDELS, P.I. S. Faber and H. Ferguson; GOODS-{\it Herschel}, P.I. D. Elbaz; CANDELS-{\it Herschel}, P.I. M. Dickinson). We do not include GOODS-N as, at the time of the writing of this paper, the IR catalog does not have uniquely identified optical counterparts. We also use photometric redshifts \citep[$z_{\rm phot}$;][]{dahlen2013,stefanon2017} and $M_\ast$ \citep{santini2015,stefanon2017}. The stellar masses are derived by fitting the CANDELS UV/Optical photometry in ten different ways, each fit using a different code, priors, grid sampling, and star formation histories (SFHs). The final $M_\ast$ is the median from the different fits, and it is stable against the choice of SFH and the range of metallicity, extinction, and age parameter grid sampling.
The CANDELS $z_{\rm phot}$s are the median redshift determined through five separate codes that fit templates to the UV/optical/near-IR data \citep[the technique is fully described in][]{dahlen2013}. Taking the median of several methods improves the accuracy, and comparison of $z_{\rm phot}$s with spectroscopic redshifts for a limited sample gives $\sigma = 0.03$ where $\sigma$ is the rms of $(z_{phot}-z_{spec})/(1+z_{spec})$. As we sort sources into redshift bins of $\Delta z = 0.5$, we do not expect the uncertainty on the photometric redshifts to be a dominant source of uncertainty in our results. We will be using the $z_{\rm phot}$s to help classify sources as AGN, SFGs, or Composites. 

MIPS 24\,$\mu$m and {\it Herschel} PACS and SPIRE catalogs were built following the prior-based PSF fitting method described in \citet[MIPS photometry]{gonzalez2005} and \citet[merged MIPS plus {\it Herschel} photometry]{gonzalez2010}. For additional details on the methods used for {\it Herschel} catalog building, see \citet{rawle2016}. Briefly, the algorithm uses IRAC and MIPS data to extract photometry for sources in longer wavelength data using positional priors. Deblending is not possible when sources lie closer than 75\% of the FWHM of the PSF in each band, making this value a minimum separation required to perform deblending. 
The final product of the cataloging method is a list of IRAC sources with possible counterparts in all longer wavelength data. In this sense, several IRAC sources might be identified with the same MIPS or {\it Herschel} source. This is what we call multiplicity. The multiplicity for MIPS and PACS is in more than 95\% of the cases equal to 1 (i.e., only one IRAC source is identified with a single MIPS and PACS source), but it is higher for SPIRE (on average, 6 IRAC sources are found within the FWHM of the SPIRE 250\,$\mu$m PSF). In order to identify the ``right'' IRAC counterpart for each far-IR sources, we follow the method described in Rodr\'iguez-Mu\~n\'oz et al. (2017, in prep). In practice, we choose the MIPS most probable 
counterpart as the brightest IRAC candidate. Then, we shift this methodology to longer wavelength bands. We identify the most likely PACS counterpart as the brightest source in MIPS 24\,$\mu$m among the different candidates. When MIPS is not available, we use the reddest IRAC band in which the source is detected. We note that using IRAC as a tracer of PACS emitters can lead to spurious 
identifications. For this reason, these cases are flagged to evaluate the possible impact in the results. Finally, we use the fluxes in PACS or MIPS (if PACS is not available) to find the counterparts of the SPIRE sources. The flux of each FIR source is assigned to a single IRAC counterpart. The FWHM of the PACS PSF is roughly the same as for MIPS, so the most serious concern in this work is matching to the SPIRE 250\,$\mu$m sources. We are primarily using the IR photometric catalogs to calculate galaxy counts. As a check, we remove all classifications of galaxies (as SFG, AGN, and Composites) that were done with SPIRE data (described in the following section). Our main result, the galaxy counts at cosmic noon, are unchanged, giving confidence that any misidentification of a SPIRE sources with a MIPS and IRAC counterpart is not biasing our results. 

We have also added sources from the COSMOS survey \citep{scoville2007} which are necessary to boost the bright end of the galaxy counts, due to the small survey area of CANDELS (0.22\,deg$^2$). We use the public COSMOS2015 catalog in \citet{laigle2016}, which presents multiwavelength data as well as stellar masses and photometric redshifts. The {\it Spitzer} IRAC data in this catalog originally comes from SPLASH COSMOS and S-COSMOS \citep{sanders2007} while the MIPS 24\,$\mu$m observations are described in \citet{lefloch2009}. The catalog also contains {\it Herschel} observations from the PEP guaranteed time program \citep{lutz2011} and the HERMES consortium \citep{oliver2012}. The counterpart identification and procedures for measuring stellar masses and photometric redshifts are fully described in \citet{laigle2016}.

The difficulty in matching MIPS, PACS, and SPIRE sources with their IRAC counterparts underscores the improvements that will be made by using MIRI color selection to identify AGN host galaxies, since the much smaller PSF ($<1''$ for all filters) and smaller spectral range used will obviate the need for counterpart identification for robust color diagnostics. 

\section{IR identification of AGN and Composites}
\label{fake}
To identify SFGs, Composites, and AGN, we build on the color techniques in \citet{kirkpatrick2013,kirkpatrick2015} that sample the full IR SED. At $z\sim1-2$, the color $S_8/S_{3.6}$ separates sources with a strong stellar bump, present in SFGs, from those with hot torus emission, found in AGN. Composites span a range in this color, depending on the ratio of relative strengths of the AGN and host galaxy emission and the amount of obscuration of the AGN due to dust.\footnote{In fact, heavily obscured AGN such as Mrk 231, NGC 1068, the Circinus galaxy, and IRAS 08572+3915 have SEDs that drop rapidly from 10\,$\mu$m toward shorter wavelengths and will show the near IR stellar spectral peak characteristic of SFGs. Hereafter, we refer to `AGN' with the understanding that the samples discussed may suffer from incompleteness of sources like these. This issue is discussed further in Section 4.1.} $S_{100}$ and $S_{250}$ trace the peak of the IR SED, which is generally dominated by the cold dust in the diffuse ISM. $S_{24}$ traces the PAH emission in SFGs or the warm dust emission heated by the AGN. Then, the color $S_{250}/S_{24}$ or $S_{100}/S_{24}$ will measure the relative amounts cold emission to warm dust or PAH emission, and this ratio is markedly higher in SFGs. However, significant scatter is introduced into color selection by redshift, since $S_{24}$ will move over different PAH features and silicate absorption at 9.7\,$\mu$m, changing where SFGs lie in color space. We can more robustly identify SFGs, AGN, and Composites if we introduce a redshift criterion.

The color diagnostics ($S_{250}/S_{24}$ vs. $S_8/S_{3.6}$ and $S_{100}/S_{24}$ vs. $S_8/S_{3.6}$) were calibrated with a sample of 343 galaxies with {\it Spitzer} IRS spectroscopy and $S_{24}>0.1\,$mJy spanning the range $z\sim0.5-4$ and $M_\ast >10^{10}\,M_\odot$. This sample is fully described in \citet{kirkpatrick2012}, \citet{sajina2012}, and \citet{kirkpatrick2015}. We identified SFGs, Composites, and AGN through spectral decomposition, where we fit the mid-IR spectrum ($5-18\,\mu$m restframe)
with a model consisting of PAH features for star formation, a power-law continuum for the AGN, and extinction. We then quantified
the AGN emission, \fAGN, as the fraction of mid-IR luminosity ($5-15\,\mu$m) due to the power-law
continuum. We define three classes of galaxies based on \fAGN, and we also report the fraction of MIR luminosity solely due to emission from the PAH features in the $5-15\,\mu$m range: (1) SFGs are dominated by
PAH emission ($\fAGN<0.2$, $L_{\rm PAH}/L_{\rm MIR} > 0.6$); (2) AGN have negligible PAH emission ($\fAGN>0.8$, $L_{\rm PAH}/L_{\rm MIR} < 0.15$);
(3) Composites have a mix of PAH and continuum emission ($\fAGN=0.2-0.8$, $L_{\rm PAH}/L_{\rm MIR} = 0.15 - 0.6$). We note that below, we will redefine these thresholds for color selection.
We relate the mid-IR classification to the full IR SED by creating
empirical templates using data from {\it Spitzer} and {\it Herschel}. We sort sources into subsamples based
on \fAGN, and after normalization, determine the median $L_\nu$ in differential bin sizes of $\lambda$ \citep{kirkpatrick2012,kirkpatrick2015}.
The \citet{kirkpatrick2015} SEDs are the first comprehensive public library of IR templates specifically
designed for high redshift galaxies that account for AGN emission.

We create a redshift dependent color diagnostic through use of the empirical MIR-based template Library from \citet{kirkpatrick2015}.\footnote{There are many AGN templates in the literature. In the $1-20\,\mu$m (rest wavelength) range critical for most of our color sorting the AGN templates agree well \citep{lyu2017}. At wavelengths longer than 20\,$\mu$m, there is considerable divergence; fortunately for our goals, the star forming output is so dominant by 100 and 250\,$\mu$m that the range of possibilities for AGN output has little effect on our results.} We use a template library because our spectroscopic sample of 343 sources is not large enough to separate sources into multiple $z$ bins. The MIR-based Library contains 11 templates created from our spectroscopic sources that demonstrate the change in IR spectral shape as the contribution of the AGN to the mid-IR luminosity increases, in steps of $\Delta\fAGN=0.1$. We randomly redshift each template 500 times, uniformly sampling a redshift distribution from $z=0.75-2.25$. We convolve each redshifted template with the observed frame IRAC, MIPS, PACS, and SPIRE transmission filters to create photometry, and then we resample the photometry within the template uncertainties at that particular wavelength, following a Gaussian distribution. We now have a catalog of 5500 synthetic galaxies, where we know the intrinsic AGN contribution, that represent the scatter in colorspace of real galaxies.

Next, we create color diagrams in redshift bins of $z=0.75-1.25$, $z=1.25-1.75$, and $z=1.75-2.25$. Beyond this redshift, it becomes too difficult to reliably separate Composites from SFGs with these colors. Because only a fraction of CANDELS and COSMOS sources have a SPIRE or PACS detection, we also create a color diagnostic using the colors $S_{24}/S_{8}$ vs. $S_8/S_{3.6}$, although this is slightly less accurate. In each redshift bin, we divide the color space into regions of $0.2\times0.2$\,dex and calculate the average \fAGN\ and standard deviation, $\sigma_{\rm AGN}$, of all the synthetic galaxies that lie in that region. In the Appendix, we show our three diagnostics: $S_{250}/S_{24}$ v. $S_8/S_{3.6}$ (used when a galaxy has the appropriate photometry, as it is the most complete at selecting Composite galaxies), $S_{100}/S_{24}$ v. $S_8/S_{3.6}$ (used when a galaxy does not have a 250\,$\mu$m detection), and $S_{24}/S_8$ v. $S_8/S_{3.6}$ (used for all galaxies without a longer wavelength detection). 

Our color diagnostics assign sources an \fAGN\ in bins of $\Delta\fAGN=0.1$, but the $\sigma_{\rm AGN}$ of each region is often larger than this (see the Appendix for a visual representation). Therefore, it is more accurate to broadly group sources as SFGs, Composites, and AGN. We determine how to group sources by comparing the \fAGN\ assigned to each synthetic galaxy by the three different color diagnostics. There is a one-to-one correlation between \fAGN(250\,\micron), \fAGN(100\,\micron), and \fAGN(24\,\micron), with a scatter of $\sigma=0.15$. Accordingly, we classify as SFGs sources with 
$\fAGN<0.30$, while the AGN have $\fAGN >0.70$, and Composites are everything in between. 

We assess the completeness and reliability of our color technique by determining how many of our synthetic galaxies are correctly identified as SFGs, Composites, and AGN in each diagnostic, and we list the completeness and reliability in Table \ref{complete24}. In the following definitions, we use $N_{\rm input}$ to represent the total number of intrinsic objects (so $N_{\rm AGN,input}$ is number of synthetic galaxies that are intrinsically AGN) and $n_{\rm sel}$ to represent the number of objects recovered by our color criteria (so $n_{\rm AGN,sel}$ is the number of intrinsic AGN that our color selection identifies as AGN). Completeness is defined as the fraction of AGN (for example) selected: $n_{\rm AGN,sel}/N_{\rm AGN,input}$. Reliability is the fraction of all the sources selected by the diagnostic as AGN (for example) that actually are, intrinsically, AGN: $n_{\rm AGN,sel}/n_{\rm all,sel}$. The lower completeness and reliability of the Composites and SFGs is due to these sources being more easily confused with each other when relying on the limited SED coverage (particularly of the mid-IR) provided by 8.0, 24, 100, and 250\,$\mu$m. By adding more bands, MIRI will allow for a more nuanced measurement of the strength of the PAH emission compared with continuum and stellar bump emission. It is also important to note that we are missing AGN with extreme obscuration, whose IR colors could mimic those of SFGs. We discuss this issue more fully in Section 4.1.

We assign each CANDELS or COSMOS source with $z=0.75-2.25$ an \fAGN\ and associated uncertainty ($\sigma_{\rm AGN}$) and then broadly group sources into SFGs, Composites, and AGN. 
Overall, from CANDELS (COSMOS), 534 (6426) sources have been classified with $S_{250}/S_{24}\,v.\,S_8/S_{3.6}$, 864 (175) with $S_{100}/S_{24}\,v.\,S_8/S_{3.6}$, and 871 (5360) with $S_{24}/S_{8}\,v.\,S_8/S_{3.6}$. From CANDELS, we also fit an additional 111 sources, which lie slightly beyond the regions (within 0.2 dex) in our color classification scheme, with the \citet{kirkpatrick2015} template library to determine the classification.

\begin{deluxetable*}{lccccc}
\tablewidth{3in}
\tablecaption{Completeness (Reliability) of Redshift Dependent Color Selection\label{complete24}}
\tablehead{\colhead{Region} & \colhead{$z\sim1$} & \colhead{$z\sim1.5$}& \colhead{$z\sim2$}}
\startdata
\cutinhead{$S_{250}/S_{24}\ v.\ S_{8}/S_{3.6}$}
AGN 	& 93 (85)\%	& 92 (89)\%		& 97 (86)\% \\
Composite & 67 (67)\%	& 81 (63)\%		& 56 (66)\%	\\
SFG 	& 66 (77)\% 	& 41 (75)\%		& 64 (64)\%	\\

\cutinhead{$S_{100}/S_{24}\ v.\ S_{8}/S_{3.6}$}
AGN 	& 97 (89)\%		& 94 (89)\%		& 98 (81)\%	\\
Composite & 69 (71)\% 	& 76 (64)\%		& 42 (59)\%	\\
SFG 	& 67 (75)\%		& 46 (69)\%		& 62 (57)\%	\\

\cutinhead{$S_{24}/S_{8}\ v.\ S_{8}/S_{3.6}$}
AGN 	& 93 (85)\%		& 90 (88)\%		& 97 (83)\%	\\
Composite & 69 (65)\%	& 67 (66)\%	& 48 (82)\%	\\
SFG		& 60 (77)\%		& 61 (65)\%		& 87 (68)\%	

\enddata
\tablenotetext{}{Completeness is defined as the percentage of sources of a given intrinsic classification that are also selected by the color diagnostic. Reliability (shown in parenthesis) is the percentage of all sources classified in a given category where the intrinsic classification agrees.}
\end{deluxetable*}

\begin{figure}
\includegraphics[width=3in]{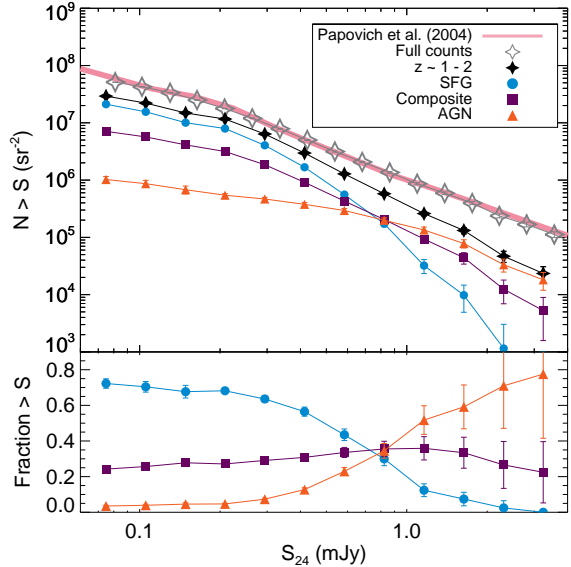}
\caption{{\it Top panel}--Cumulative 24\,$\mu$m number counts for the CANDELS and COSMOS fields. The open grey stars show the number counts of all 24\,$\mu$m detected sources, and these agree with the published number counts \citep[pink solid line]{papovich2004}. The filled stars show the 24\,$\mu$m counts from $z=0.75-2.25$. At the bright end, the lack of sources is due to the relatively small field sizes. We then show the contribution of SFGs ($\fAGN<0.3$; blue circles), Composites ($0.3\leq\fAGN<0.7$; purple squares), and AGN ($\fAGN\geq0.7$; orange triangles) to the $z\sim1-2$ number counts. {\it Bottom panel}--The percentage of each subsample above a given flux threshold. At $S_{24}>0.8\,$mJy, Composites and AGN dominate samples. Even at fainter fluxes, JWST/MIRI samples will contain $>$25\% Composites. \label{num_counts}}
\end{figure}

\subsection{Galaxy and AGN Counts}
Now, we determine how traditional 24\,$\mu$m number counts break down into the SFG, Composite, and AGN categories. We only consider sources with $S_{24}>80\,\mu$Jy, which is the 80\% completeness limit \citep{gonzalez2005}. We measure directly the EGS, COSMOS, GOODS-S, and UDS field sizes covered by our sources. We show the total CANDELS+COSMOS 24\,$\mu$m number counts as the open grey stars in Figure \ref{num_counts}, and these counts are in agreement with the counts from \citet{papovich2004}. We plot the 24\,$\mu$m counts at cosmic noon ($z=0.75-2.25$) as the filled black stars. There is a disagreement with the full counts that arises from applying the redshift cut, and this chiefly affects the bright end ($S_{24} > 1\,$mJy), which is where AGN will dominate the counts \citep{kirkpatrick2013}. The lack of bright sources is a result of the small field sizes of CANDELS (0.22\,deg$^2$) and COSMOS (2\,deg$^2$). We show how the cosmic noon counts break down into SFGs (blue), Composites (purple), and AGN (orange). We have calculated uncertainties on the counts using a Monte Carlo technique, where we vary the \fAGN\ for each source within its associated uncertainty and recount sources. We follow this procedure 1000 times. The counts in Figure \ref{num_counts} represent the mean from the Monte Carlo simulations, and the error bars are standard deviation from the Monte Carlo trials and the standard Poisson errors, summed in quadrature. 

Below 0.8\,mJy, SFGs dominate the counts, but AGN become more prevalent with increasing brightness. In the bottom panel of Figure \ref{num_counts}, we show the percentage of sources above a given flux threshold. We find that AGN contribute $\sim10\%$ at 0.3\,mJy and increase to $\sim80\%$ at 2\,mJy, in good agreement with measurements in \citet{brand2006} in the Bo\"otes field. Although AGN are frequently assumed not to be abundant in fainter IR samples, the presence of AGN hosts at $S_{24}<100\,\mu$Jy was also seen in a small {\it Spitzer}/IRS spectroscopic sample of lensed galaxies at $z\sim2$, where the authors found that 30\% of the sample had IR AGN signatures and 40\% had X-ray AGN signatures \citep{rigby2008}. The Composites comprise $>$25\% of a sample down to the faintest flux threshold at 63\% completeness, which we determined by applying the completeness estimates listed in Table \ref{complete24} to the number of sources classified with each method.  Then, at least 25\% of a JWST/MIRI sample will be Composite galaxies, providing a rich data set for probing the AGN/star formation connection at cosmic noon. 

\section{JWST Color Selection}
\label{sec:colors}
\begin{figure*}
\centering
\includegraphics[width=3in]{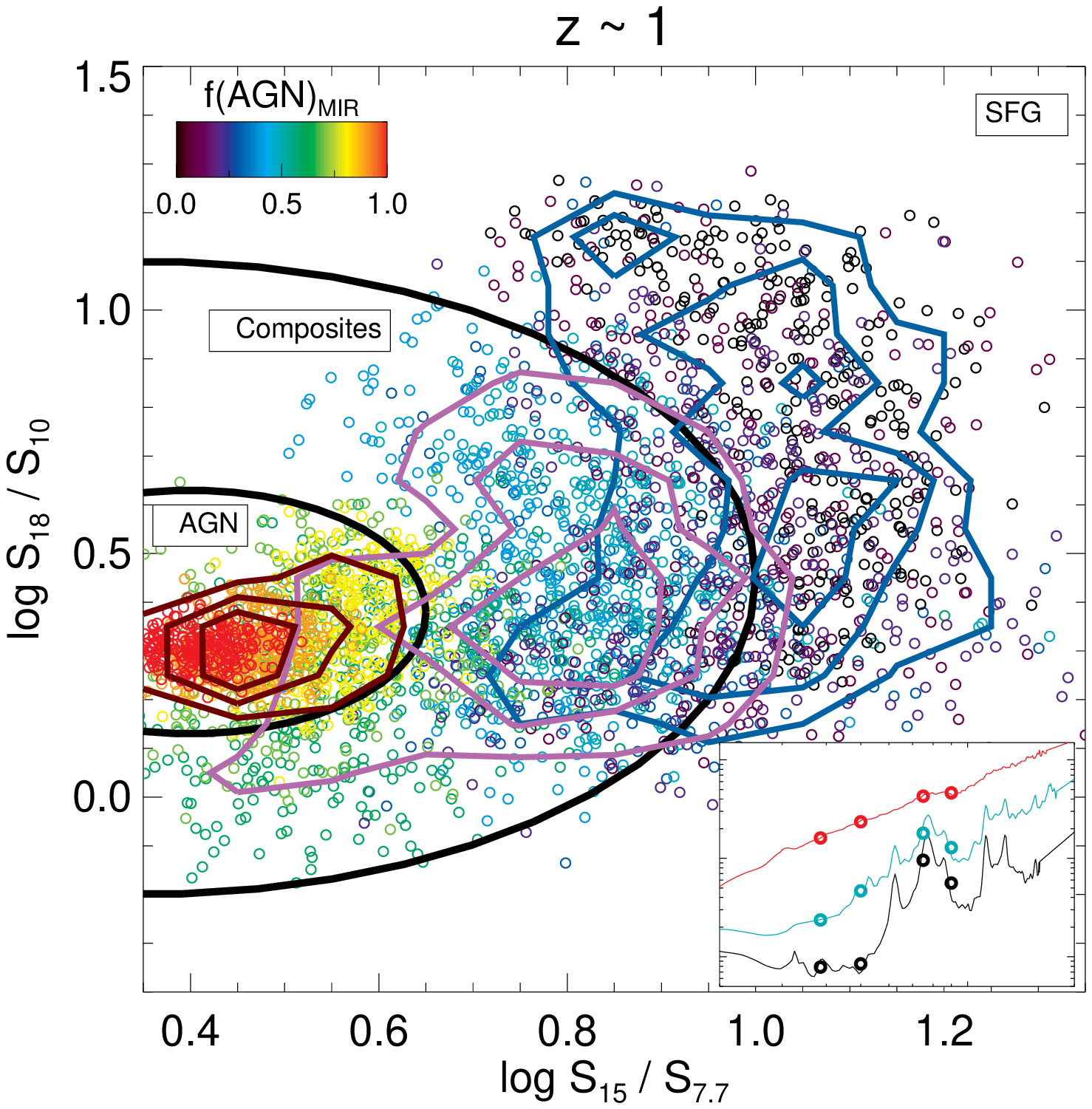}
\includegraphics[width=3in]{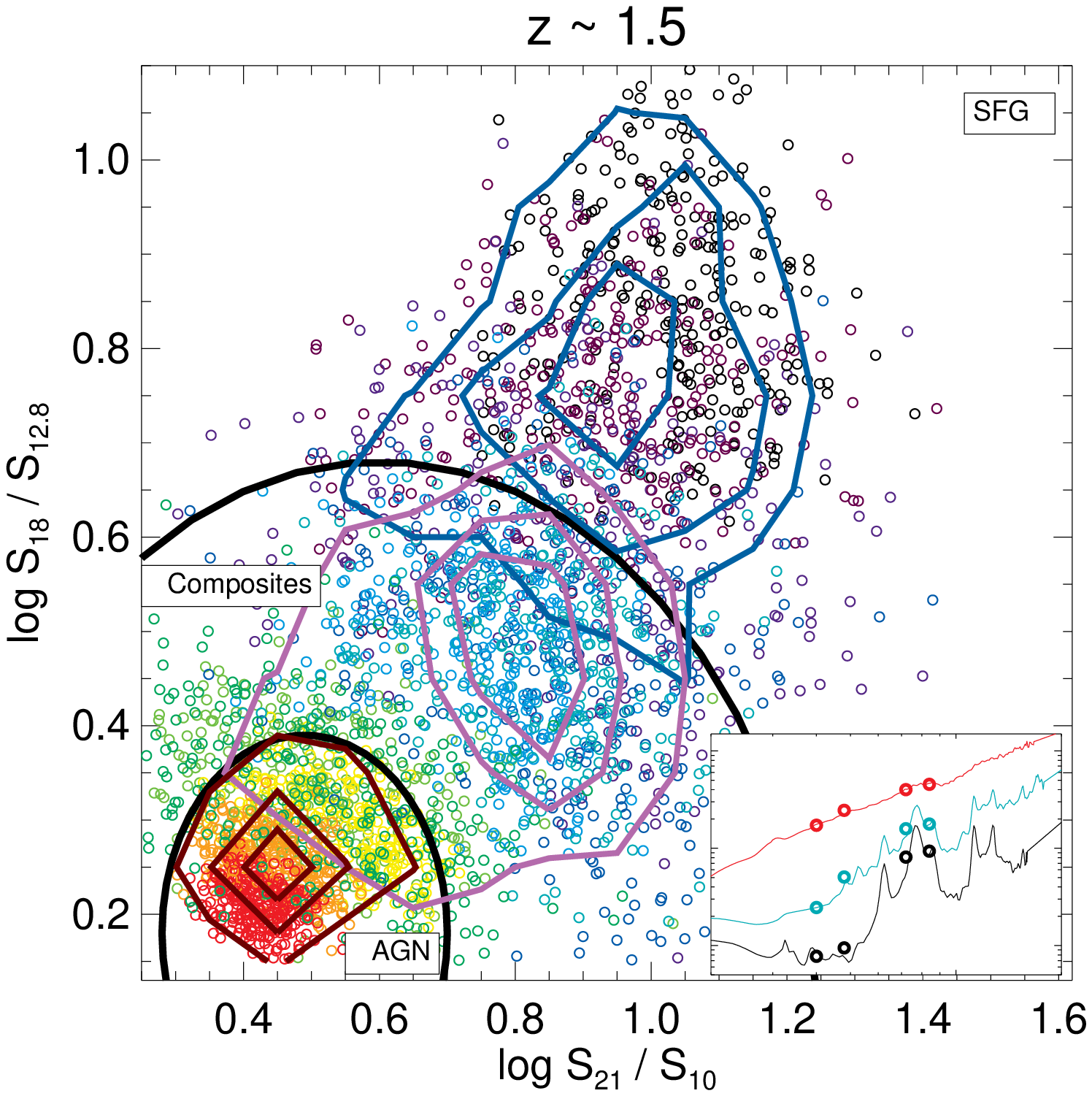}
\includegraphics[width=3in]{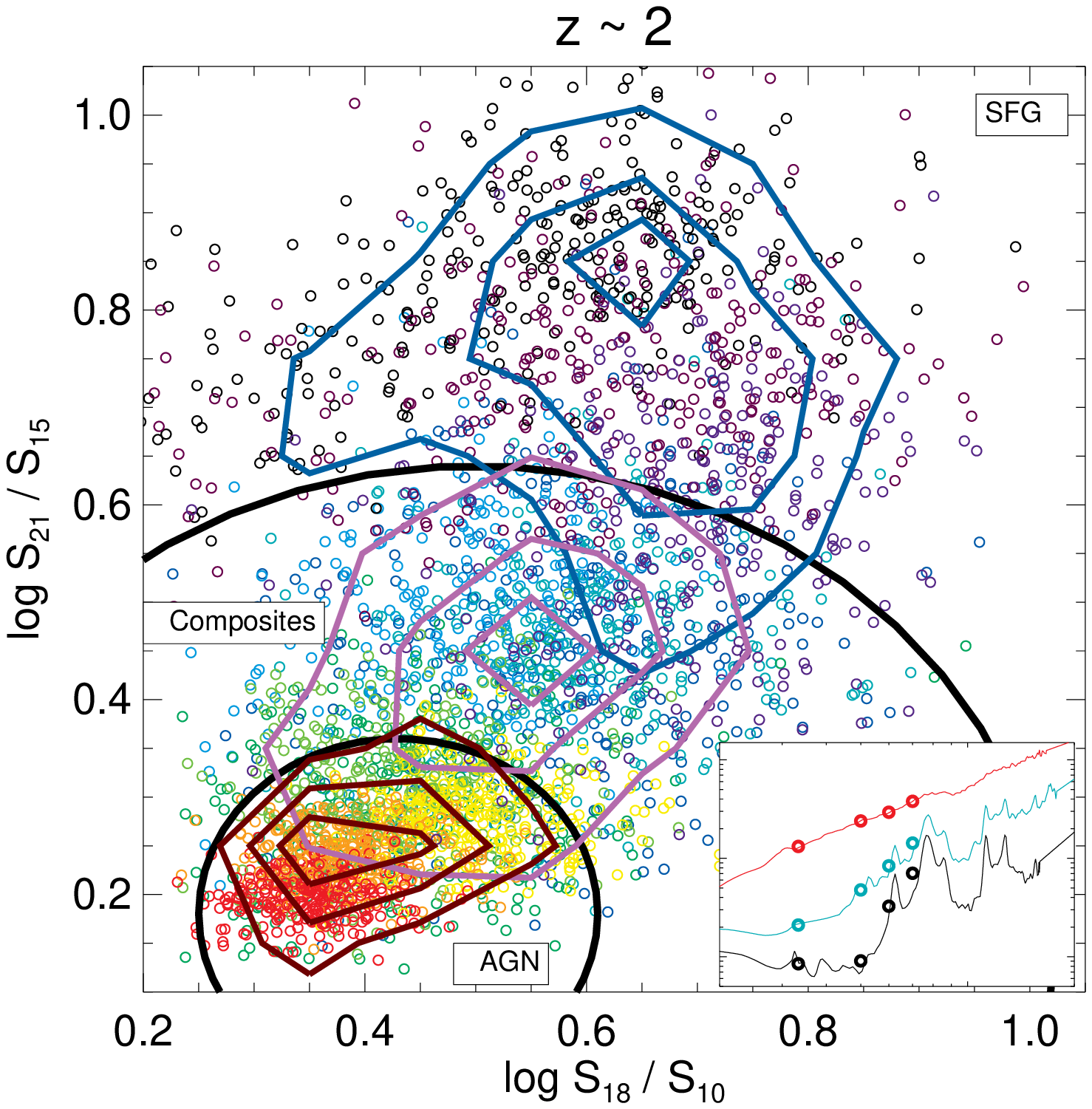}
\caption{The optical MIRI color combination for separating Composites, AGN, and SFG from $z=0.75-1.25$ ({\it Top Left}), $z=1.25-1.75$ ({\it Top Right}), and $z=1.75-2.25$ ({\it Bottom}). We show the synthetic galaxies (shaded according to \fAGN) created from the \citet{kirkpatrick2015} library used to determine the best AGN and Composite selection regions (black lines), based on completeness and reliability. We also overplot the contours of all the synthetic galaxies classified as SFG (blue lines), Composites (purple lines), and AGN (maroon lines) to allow easier viewing of where each category predominantly lies. In the bottom right corner of each diagram, we demonstrate where the photometry filters fall on an SFG (black), Composite (blue), and AGN (red) template at $z=1,1.5,2$. \label{colors}}
\end{figure*}

Color selection is a powerful technique for identifying likely AGN, Composites, and SFGs. We have done an exhaustive search to identify the best MIRI filter combinations for separating galaxies into these three classes at cosmic noon by creating synthetic photometry in the JWST/MIRI filters from the \citet{kirkpatrick2015} MIR based Library following the Monte Carlo technique outlined in Section \ref{fake}. 
As many JWST/MIRI observations will be carried out in fields with available photometric redshifts, or in parallel with NIRcam and NIRspec observations, we include redshift information in our color diagnostics to improve reliability and completeness. We identify three diagnostics covering the ranges $z\sim1$ ($z=0.75-1.25$), $z\sim1.5$ ($z=1.25-1.75$), and $z\sim2$ ($z=1.75-2.25$). These three diagnostics, shown in Figure \ref{colors}, are different combinations of the $S_{21}, S_{18}, S_{15}, S_{12.8}, S_{10},$ and $S_{7.7}$ filters, which cover the 6.2 and 7.7\,$\mu$m PAH complexes and the $3-5\,\mu$m stellar minimum at these redshifts. 

We present two methods for separating SFGs, Composites, and AGN. First, we have determined the optimal AGN, Composite, and SFG regions, labeled in Figure \ref{colors}. The boundaries of each region are circles, with AGN lying inside the inner circle, SFGs lying outside the outer circle, and Composites lying in between.

The $z\sim1$ boundaries are
\begin{align}
&inner{\rm :}\ (\log \frac{S_{15}}{S_{7.7}}-0.40)^2 + (\log \frac{S_{18}}{S_{10}}-0.38)^2 = 0.25^2\nonumber\\
&outer{\rm :}\ (\log \frac{S_{15}}{S_{7.7}}-0.35)^2 + (\log \frac{S_{18}}{S_{10}}-0.45)^2 = 0.65^2
\end{align}

The $z\sim1.5$ boundaries are
\begin{align}
&inner{\rm :}\ (\log \frac{S_{21}}{S_{10}}-0.49)^2 + (\log \frac{S_{18}}{S_{12.8}}-0.18)^2 = 0.21^2\nonumber\\
&outer{\rm :}\ (\log \frac{S_{21}}{S_{10}}-0.60)^2 + (\log \frac{S_{18}}{S_{12.8}}-0.03)^2 = 0.65^2
\end{align}

The $z\sim2$ boundaries are
\begin{align}
&inner{\rm :}\ ( \log \frac{S_{18}}{S_{10}}-0.43)^2 + (\log \frac{S_{21}}{S_{15}}-0.18)^2 = 0.18^2\nonumber\\
&outer{\rm :}\ ( \log \frac{S_{18}}{S_{10}}-0.50)^2 + (\log \frac{S_{21}}{S_{15}}-0.12)^2 = 0.52^2
\end{align}

These regions are useful for broadly classifying large numbers of sources or identifying targets for follow-up observations. We use these regions to assess the reliability and completeness of our color diagnostic, where again, we classify all synthetic sources as SFGs when $\fAGN < 0.3$, Composites where $0.3 \leq \fAGN<0.7$, and AGN when $0.7 \geq \fAGN$. Table \ref{table:colors} lists these values for all three redshift regimes. Comparison with Table \ref{complete24} shows an improvement over what we were able to reliably classify with the {\it Herschel} and {\it Spitzer} diagnostics, particularly for separating Composites from SFGs. The spacing of the MIRI filters allows us to sensitively trace the strength of the PAH features relative to the stellar minimum, where the proportionate amount of PAH emission will be lower for Composite galaxies as the power-law emission from the AGN begins to outshine the stellar minimum (see the insets in Figure \ref{colors} for a visual guide).

Perhaps, instead of broad classifications, the reader would rather have an estimate of \fAGN. Without mid-IR spectroscopy, robust decomposition into an AGN and star forming component still is not feasible, even with 6 photometry filters. However, we have determined how to linearly combine the colors in each redshift regime in order to estimate \fAGN, and we also measure the standard deviation ($\sigma_{\rm AGN}$) of the residuals when each equation is applied to our synthetic sources so that the reader has a measure of the uncertainty. At $z\sim1$
\begin{equation}
\fAGN =-0.97 \times (\log \frac{S_{15}}{S_{7.7}}) -0.10\times (\log \frac{S_{18}}{S_{10}})+1.29
\end{equation}
and $\sigma_{\rm AGN} = 0.15$.

At $z\sim1.5$:
\begin{equation}
\fAGN = -0.56\times (\log \frac{S_{21}}{S_{10}}) -0.85 \times (\log \frac{S_{18}}{S_{12.8}}) +1.29
\end{equation}
and $\sigma_{\rm AGN} = 0.13$.

At $z\sim2$:
\begin{equation}
\fAGN = -0.55 \times (\log \frac{S_{18}}{S_{10}}) -1.01\times(\log \frac{S_{21}}{S_{15}}) +1.25
\end{equation}
and $\sigma_{\rm AGN}=0.16$.


\subsection{Mid-IR concerns: Metallicity and Obscuration}
\label{metal}
At cosmic noon, the bulk of the star formation is occurring in massive, dusty galaxies with $M_\ast>10^{10}\,M_\odot$ \citep[e.g.][]{murphy2011,madau2014,pannella2015}, which is the type of galaxies that our MIRI diagnostics were created from \citep{kirkpatrick2012,sajina2012,kirkpatrick2015}. For studying the AGN-star formation connection, we expect these types of galaxies to form the most appealing targets. Nevertheless, the sensitivity of JWST/MIRI will enable studies of lower mass galaxies, which tend to have lower metallicities \citep[and references therein]{ma2015}.
Decreasing gas phase metallicities have been linked with decreasing PAH strengths \citep[e.g.][]{engelbracht2008,sandstrom2012,shivaei2017}, which is a source of concern since we are effectively detecting AGN hosts based on the strength of PAH features compared with the stellar minimum at $3-5\,\mu$m. \citet{shipley2016} find that below $Z<0.7\,Z_\odot$, PAH emission no longer scales linearly with $L_{\rm IR}$, which based on the mass-metallicity relation, could be a source of concern for contamination of our Composite regions at $M_\ast < 3\times 10^{9} M_\odot$, up to $z\sim2.3$ \citep{erb2006,zahid2013,sanders2015}. Recently, using the MOSDEF optical spectroscopic survey, \citet{shivaei2017} found that at $z\sim2$, $L_{\rm 7.7}/L_{\rm IR}$ is lower for galaxies with $M_\ast < 10^{10}\,M_\odot$ with a behavior similar to that seen for local galaxies \citep{engelbracht2008,shipley2016}. 

\begin{deluxetable}{lccc}[ht!]
\tablecaption{Completeness (Reliability) of MIRI color selection \label{table:colors}}
\tablehead{\colhead{Region} & \colhead{$z\sim1$} & \colhead{$z\sim1.5$}& \colhead{$z\sim2$}}
\startdata
AGN 		& 87 (90)\% 	& 87 (89)\%		& 87 (80)\% \\
Composite 	& 77 (72)\%		& 79 (74)\%		& 71 (71)\%	\\
SFG 		& 76 (81)\% 	& 81 (86)\%		& 79 (89)\%	
\enddata
\end{deluxetable}

At $z\sim2$, a main sequence galaxy with $M_\ast = 10^{10}\,M_\odot$ will have a SFR of $\sim45\,M_\odot$/yr \citep{rosario2013}. At $z\sim2$, 21\,$\mu$m is tracing the 7.7\,$\mu$m PAH feature, so applying Equation 11 from \citet{shipley2016} for this SFR gives $S_{21}\approx30\,\mu$Jy, which is achievable in 7 minutes for a 10$\sigma$ detection. An hour of integration time at 21\,$\mu$m will produce 10$\sigma$ detections of galaxies at roughly 8\,$\mu$Jy, corresponding to $M_\ast \sim 3\times10^9\,M_\odot$, which is well below the threshold where we expect low metallicity galaxies might contaminate the Composite regime. As such, our color diagnostics may require recalibration for low metallicity galaxies when using observations below $S_{21}\lesssim30\,\mu$Jy.

\begin{figure}
\includegraphics[width=3in]{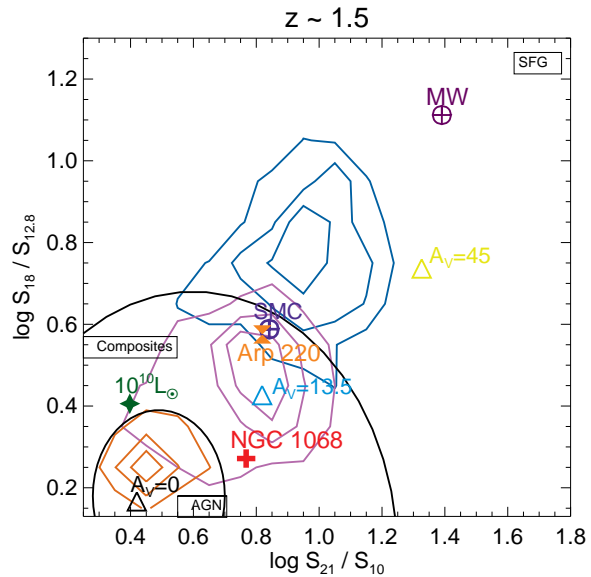}
\caption{We use our $z\sim1.5$ MIRI diagnostic to explore where sources with different luminosities, metallicities, and obscurations than the \citet{kirkpatrick2015} library will lie in colorspace; we show the distribution of our synthetic SFGs (blue), Composites (purple), and AGN (orange) as well as the black lines marking the AGN, Composite, and SFG regions. We use the \citet{draine2007} library to calculate where an SMC galaxy (dark purple circle with cross) and MW galaxy with a lower PAH fraction (purple circle with cross) will lie. The SMC galaxy lies in the Composite region, as well as a galaxy with $L_{\rm IR}=10^{10}\,L_\odot$ (dark green star) from the \citet{rieke2009} library. Our diagnostics were calibrated using galaxies with $L_{\rm IR}>10^{11}\,L_\odot$ and $M_\ast>10^{10}\,M_\odot$, so they should be applied with caution to lower mass, lower luminosity objects. We also look at what effect obscuration may have on our ability to detect AGN using local obscured ULIRG Arp 220 (orange bowtie) and Compton Thick AGN NGC 1068 (red cross), both of which lie in the Composite Region. We also use the AGN template library (triangles) of \citet{siebenmorgen2015} to show how obscuration in the torus (measured in $A_V$) will push AGN into the Composite and SFG region.\label{others}}
\end{figure}

As a visual check, we demonstrate in Figure \ref{others} where SFGs with different PAH strengths will lie in our $z\sim1.5$ diagnostic. To accomplish this, we use the Small Magellanic Cloud (SMC) dust model (PAH fraction $q_{\rm PAH}=0.10\%$) and a Milky Way dust model with $q_{\rm PAH}=0.47\%$ from \citet{draine2007}, which is included to show where a galaxy with a low SFR will lie. The \citet{draine2007} models are also parameterized in terms of the strength of the radiation field, $U_{\rm min}$ and $U_{\rm max}$. We set these values to $U_{\rm min}=1$ and $U_{\rm max} = 1e5$, although these parameters have little effect on the final colors. Also, we note that we add in a stellar blackbody with $T=5000\,$K to complete the near-IR portion of the spectrum. Even with a low PAH fraction, the Milky Way template still lies in our SFG region, while the SMC template lies directly on the Composite/SFG border. Haro 11, another well studied low metallicity galaxy \citep[$Z=1/3\,Z_\odot$,][]{james2013} in the nearby Universe has nearly identical MIRI colors as our plotted SMC data point, further confirming that low metallicity galaxies will likely lie around the Composite/SFG border. The reason is that even though low metallicity galaxies have diminished PAH features, they still have a deep and broad stellar minimum at $3-5\,\mu$m \citep{lyu2016}, unlike Composites which begin to exhibit the warmer dust characteristic of the AGN torus. We also plot the template from \citet{rieke2009} which corresponds to an $L_{\rm IR}=10^{10}\,L_{\rm IR}$, as this is an order of magnitude less luminous than the \citet{kirkpatrick2015} library. A galaxy of this luminosity also lies in the Composite region, although it is away from the locus of our Composite galaxies (purple distribution). 

We caution the reader to be prudent when classifying galaxies as Composites, particularly low mass sources that lie near the Composite and SFG border.
If stellar masses of MIRI samples are known (possibly through NIRcam observations), low mass galaxies that lie in our Composite regions provide excellent targets for follow-up spectroscopy observations, to distinguish between AGN or metallicity as the underlying cause of the diminished PAH emission.

The other prominent concern in a mid-IR diagnostic is how obscuration can affect the detection of AGN. Our template library was built assuming the AGN can be represented as a power law, and we empirically measure the power law component to have an average slope of $F_\nu\propto\lambda^{1.5}$, but individual sources will show a range of slopes, and a range of dust obscurations. The AGN templates in the \citet{kirkpatrick2015} library are derived from AGN where 75\% of the sample are also detected in the X-ray, implying that they are largely unobscured. Of the Composite sources in \citet{kirkpatrick2015}, only 35\% are X-ray detected, indicating that they contain more heavily obscured AGN. We now explore the effects of dust obscuration by examining where different galaxies will lie in the $z\sim1.5$ colorspace (Figure \ref{others}).

Arp 220 (orange bowtie) is a local Ultra Luminous Infrared Galaxy (ULIRG) that is heavily dust-obscured and may host an AGN \citep{veilleux2009,teng2015}. Its position near the SMC and at the edge of the Composite region indicates another possible ambiguity, that the aromatic bands tend to be suppressed in the most luminous and compact infrared galaxies. How many such objects exist at cosmic noon is not well quantified, as most galaxies of the same luminosity as local ULIRGs ($L_{\rm IR}>10^{12}\,L_\odot$) have extended ISMs \citep{papovich2009,younger2009,finklestein2011,rujopakarn2011,ivison2012,rujopakarn2016}. NGC 1068 (red cross) is an archetypal local Compton thick Seyfert II AGN. Despite its extreme obscuration, it lies securely in our Composite region, close to the AGN boundary. 

We also use the AGN library of \citet{siebenmorgen2015} to examine what extinction conditions would push an AGN into our SFG region. These AGN templates are calculated assuming the AGN IR emission arises from 2-phase dust region consisting of a torus and disk, a torus radius $R$, viewing angle, and cloud filling factor. The optical depth of the clouds in the torus is primarily what causes the AGN to move into the Composite and SFG regions, so we hold all other parameters fixed (for reference, we use the model with viewing angle$
 =67^\circ, R=1545\times10^{15}\,{\rm cm}, A_d=300,V_c=77.7\%$). This model is a pure AGN, with no star formation, but when the optical depth of the torus is $A_V=13.5$ (blue triangle), 
 the AGN model lies in our Composite region, and when $A_V=45$ (yellow triangle), the AGN lies in the SFG region. Detecting such an obscured AGN at other wavelengths would also be 
extremely challenging, and identifying complete samples of true Type II obscured AGN remains an unsolved problem. \citet{delmoro2016} find that 30\% of mid-IR luminous quasars at $z\sim1-3$ in the GOODS-S field are not detected in the {\it Chandra} 6 Ms data. Of those that are detected, $>65\%$ are Compton thick. 
Beyond these estimates, it is difficult to say how many heavily obscured AGN there are that would not be selected as such in the X-ray or the mid-IR. Identifying these very obscured AGN will require detailed SED modeling using a full suite of NIRcam+MIRI observations, which is beyond the scope of this paper.

\subsection{AGN contributions in individual bands}
If we have a good understanding of the typical full IR SED of high redshift galaxies, as well as the scatter in the population, then a single photometric point can be used in conjunction with representative templates to estimate $L_{\rm IR}$ and star formation rates (SFRs). Since PAH molecules are illuminated by the UV/optical photons from young stars, they are a natural SFR indicator and have been extensively used in the literature to probe SFR and $L_{\rm IR}$ \citep{peeters2004,brandl2006,pope2008,battisti2015,shipley2016}. 

Given the coverage of the MIRI filters, we will now examine how an AGN can affect the 7.7\,$\mu$m PAH feature for the \citet{kirkpatrick2015} templates used in this work, as any AGN contribution will need to be corrected for before converting a PAH luminosity to a SFR. We remind the reader that for these templates, the AGN component is represented as a power-law with a slope of $F_\nu \propto\lambda^{1.5}$.
We measure the intrinsic $L_{7.7}$ of each template using PAHFIT \citep{smith2007}.  
Then, we measure $L_{\rm MIRI}$, which is the photometry of the template through the following MIRI filters at the given redshifts:
\begin{align}
\label{redarr}
z=0,&\ 7.7\,\micron \nonumber \\
z=0.95,&\ 15.0\,\micron \nonumber\\
z=1.34,&\ 18.0\,\micron \nonumber\\
z=1.73,&\ 21.0\,\micron \nonumber\\
z=2.31,&\ 25.5\,\micron
\end{align}
The redshifts mark where the rest frame central wavelength of each filter is 7.7\,$\mu$m.

\begin{figure}
\includegraphics[width=3.0in]{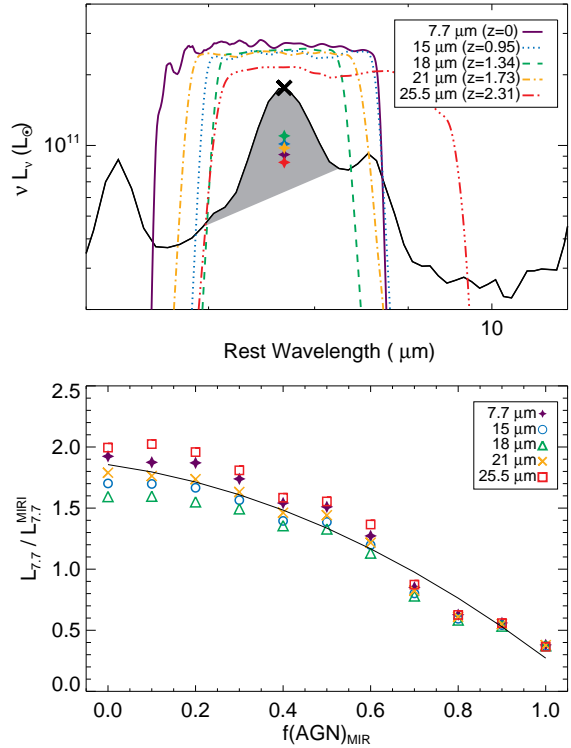}
\caption{{\it Top panel}--We demonstrate, using the MIR0.0 template from \citet{kirkpatrick2015}, how much of the 7.7\,$\mu$m feature each MIRI filter covers at the redshifts where the central wavelength of each filter is 7.7\,$\mu$m (listed in the legend). The grey shaded region indicates which part of the spectrum is integrated to calculate $L_{7.7}$ (black cross), while the filled stars show the photometry (in units of $\nu L_\nu$) measured through each of the MIRI filters. The photometry is lower because it includes more of the spectrum at lower luminosity.
{\it Bottom panel}--We show the ratio of the MIRI photometry ($L_{\rm MIRI}$) measured through different MIRI filters depending on redshift to the intrinsic $L_{7.7}$ of the PAH feature. The lower fractions with increasing \fAGN\ are due to an increased warm dust continuum due to heating by an AGN. If $L_{7.7}$ is going to be used to calculate SFRs, corrections need to be made for an AGN contribution. The black line is the empirical relationship between $L_{7.7}/L_{\rm MIRI}$ and \fAGN. \label{fig:sfrs}}
\end{figure}

In the top panel of Figure \ref{fig:sfrs}, we demonstrate how much of the 7.7\,$\mu$m feature each filter covers at the above listed redshifts. In the bottom panel, we show the relationship $L_{7.7}/L_{\rm MIRI}$ as a function of \fAGN\ for each filter at the listed redshifts. The decreasing fractions with increasing \fAGN\ are due to the increased contribution of the warm dust continuum to the measured photometry. We fit a quadratic relationship to all the points and measure
\begin{align}
 \frac{L_{7.7}}{L_{\rm MIRI}}=&(-1.09\pm0.20)\times\fAGN^2 \nonumber\\
 &-(0.50\pm0.21)\times\fAGN\nonumber\\
 &+(1.86\pm0.04)
 \end{align}
This equation, in conjunction with estimating \fAGN\ from MIRI colors, can be used for first order corrections to $L_{7.7}$ before converting to a SFR. Similarly, in \citet{kirkpatrick2015}, we demonstrated that there is a quadratic relationship between \fAGN\ and the total contribution of an AGN to $L_{\rm IR}$ that can be used to correct $L_{\rm IR}$ for AGN emission:
\begin{equation}
\label{eq:AGN}
\fTOT = 0.66 \times \fAGN^2 - 0.035\times \fAGN
\end{equation}
where \fTOT\ is the fraction of $L_{\rm IR}(8-1000\,\mu$m$)$ due to AGN heating. Then, the portion of $L_{\rm IR}$ due to star formation is $L_{\rm IR}^{\rm \scriptscriptstyle SF} = L_{\rm IR} \times(1-\fTOT)$. Once the AGN contribution is accounted for, $L_{\rm IR}$ can be converted to a SFR using standard equations \citep[e.g.,][]{murphy2011}. For a strong AGN ($\fAGN \geq 0.9$), at least 50\% of $L_{\rm IR}$ needs to be removed before converting to a SFR, and the same is true if using $7.7\,\mu$m to calculate SFR. Then, the strongest AGN will have SFRs that are overestimated by at least a factor of 2 if not properly accounted for. Of more concern is Composites, which are routinely misidentified as SFGs. For a Composite with $\fAGN=0.5$, an $L_{\rm IR}$ based SFR will be overestimated by $\sim15\%$. But, if one uses $L_{7.7}$, then the resulting SFR will be overestimated by $\sim35\%$.

\section{Discussion: Physical properties of a MIRI sample}
\begin{figure}
\includegraphics[width=3.3in]{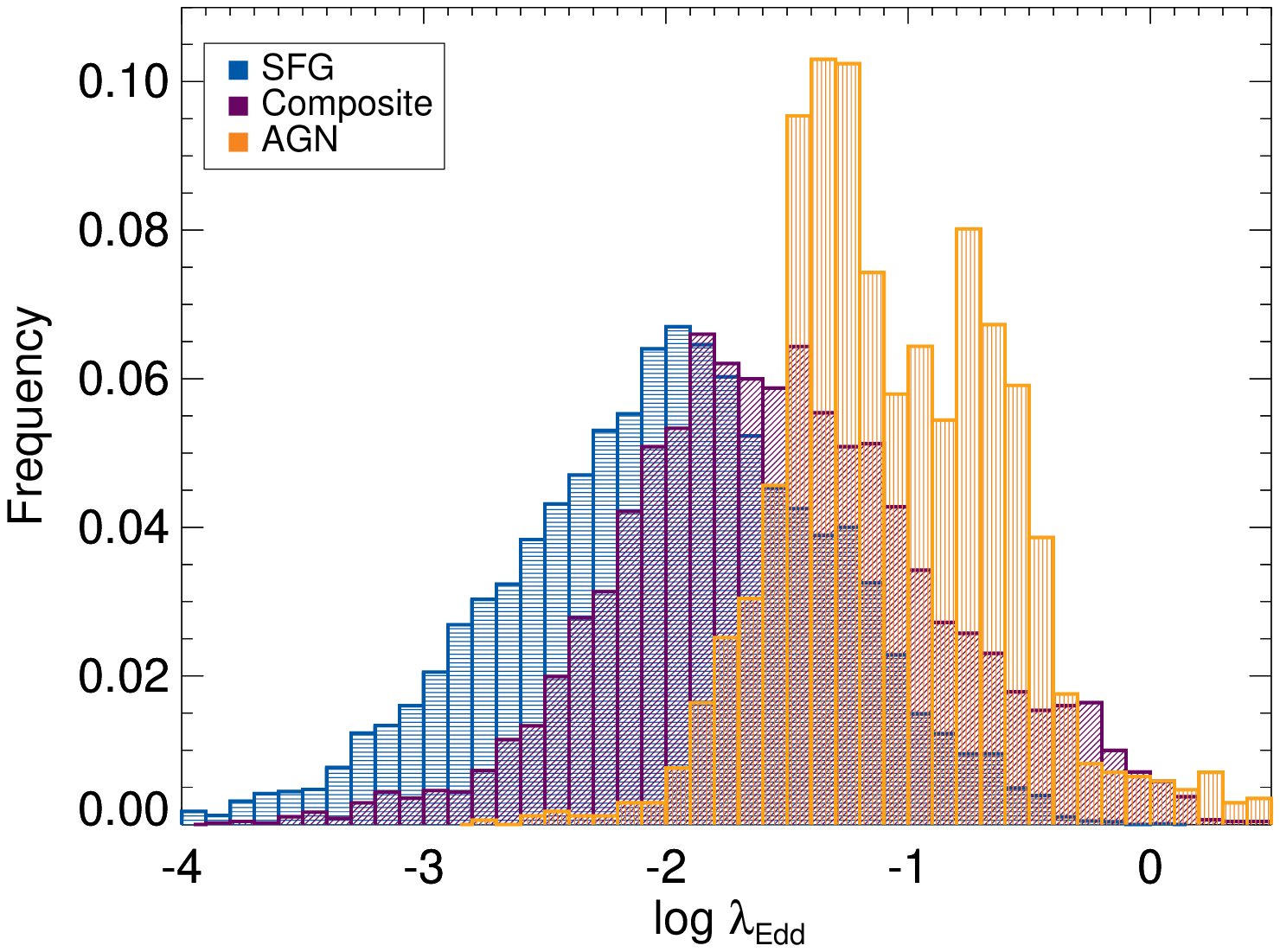}
\includegraphics[width=3.3in]{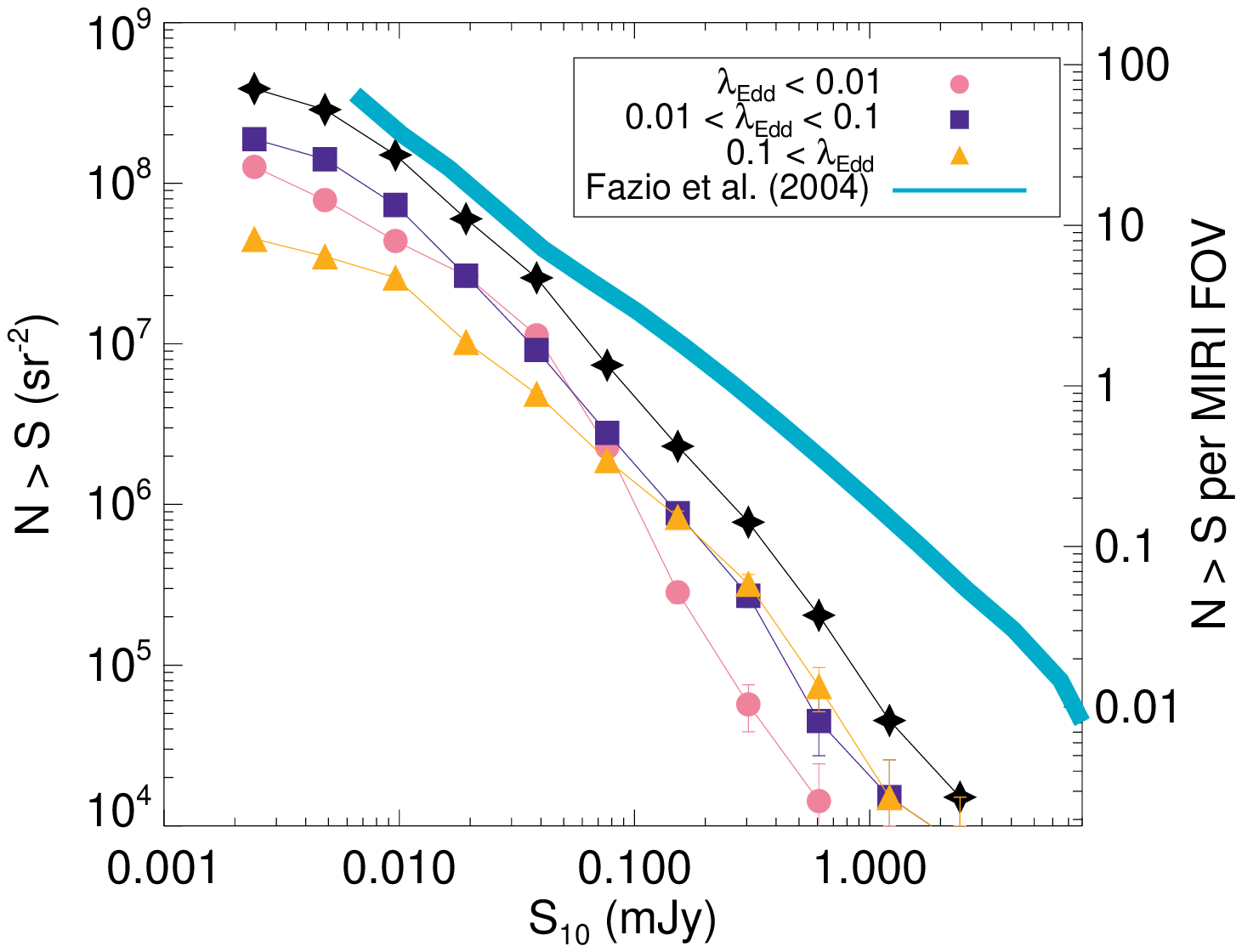}
\caption{{\it Top panel}--We calculate $\lambda_{\rm Edd}$ for CANDELS+COSMOS identified SFGs (blue), Composites (purple), and AGN (orange) by applying standard scaling relations to $M_\ast$, $\fAGN$, and $L_{\rm IR}$. We have normalized the distributions to the show relative frequencies in each category. Each category spans an overlapping range, illustrating the current limitations in understanding how observed IR dust emission relates to the accretion on a galaxy's central black hole. {\it Bottom panel}--We predict the MIRI 10\,$\mu$m number counts at $z=0.75-2.25$ and then break them into bins of $\lambda_{\rm Edd}$. We have scaled the 24\,$\mu$m emission of the CANDELS and COSMOS galaxies shown in Figure \ref{num_counts} using the appropriate template for SFGs, Composites, and AGN. Our 10\,$\mu$m counts agree with the measured 8\,$\mu$m counts in \citet[blue line]{fazio2004}. The discrepancies between the 10\,$\mu$m and 8\,$\mu$m counts can be attributed to the redshift cut. In one MIRI FOV, we will detect nearly 100 galaxies down to 2\,$\mu$Jy, and many of these will be AGN hosts. \label{FOV}}
\end{figure}

We now return to our CANDELS+COSMOS sample to investigate the physical properties of galaxies that MIRI color selection will identify as being AGN hosts. First, we illustrate the predicted number counts at cosmic noon with the MIRI 10\,$\mu$m filter, which is chosen for its sensitivity \citep[$\sim0.6\,\mu$Jy at 10$\sigma$ in $<$3\,hours;][]{glasse2015} and because we use it in all three color diagnostics. We calculate the 10\,$\mu$m flux for all CANDELS+COSMOS galaxies at $z=0.75-2.25$ and with $M_\ast > 10^8\,M_\odot$ by scaling the appropriate \citet{kirkpatrick2015} template (based on the source's \fAGN\ determined through color classification) to the available IR photometry and convolving with the 10\,$\mu$m transmission filter. By template fitting, we are also able to calculate $L_{\rm IR}$ and \fTOT. The total 10\,$\mu$m counts are plotted as the black stars in the bottom panel Figure \ref{FOV}. By including lower mass galaxies, we push below the 80\% completeness in Figure \ref{num_counts} and down to the 20\% completeness limit (corresponding to $\sim40\,\mu$Jy at 24\,$\mu$m). For reference, the 80\% completeness limit (measured at 24\,$\mu$m) corresponds to $S_{10}\sim10\,\mu$Jy. Our counts are in good agreement at the faint end with the published 8\,$\mu$m galaxy counts in \citet{fazio2004}. At the bright end, we have fewer sources due to the redshift cut we imposed and the small field sizes, similar to our 24\,$\mu$m number counts in Figure \ref{num_counts}.

\begin{figure}
\includegraphics[width=3.3in]{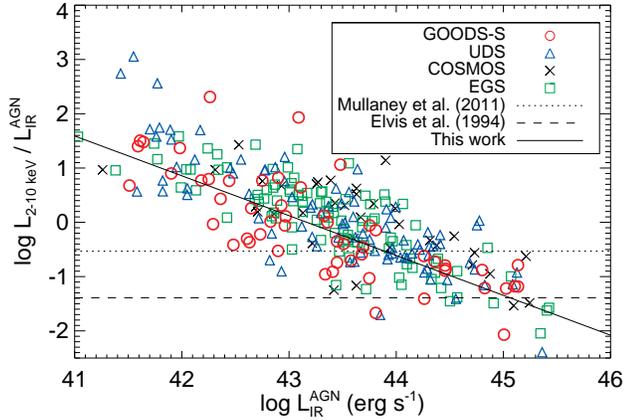}
\caption{$\frac{L_{2-10{\rm keV}}}{L_{\rm IR}^{\rm \scriptscriptstyle AGN}}$ as a function of $L_{\rm IR}^{\rm \scriptscriptstyle AGN}$ (all quantities are in erg\,s$^{-1}$) for the four CANDELS fields. We empirically measure the relationship (solid line) using only the GOODS-S data (red circles), since the is the field where the {\it Chandra} observations are complete down to $L_{2-10{\rm keV}}= 10^{42}\,$erg\,s$^{-1}$ out to $z=2$. We show the approximate conversions derived in \citet[dotted line]{mullaney2011}, using a local sample of AGN with $L_{2-10{\rm keV}}\sim10^{43}\,$erg\,s$^{-1}$, and derived in \citet[dashed line]{elvis1994} from quasars with $L_{2-10{\rm keV}}>10^{45}\,$erg\,s$^{-1}$.\label{xray}}
\end{figure}

\fAGN\ is strictly a measure of the dust heated by a AGN relative to that heated by star formation, so now we examine a more physically motivated quantity, the Eddington ratio. The Eddington ratio is defined as $\lambda_{\rm Edd}= L_{\rm bol}/L_{\rm Edd}$, where $L_{\rm bol}$ is the bolometric luminosity of the AGN and $L_{\rm Edd}$ is the Eddington luminosity. In this way, $\lambda_{\rm Edd}$ is a measure of how efficiently a black hole is accreting material. $L_{\rm bol}$ is commonly estimated from the hard X-ray luminosity, $L_{2-10{\rm keV}}$. Due to obscuration and varying depths of the {\it Chandra} catalogs in the CANDELS fields, we do not have $L_{2-10{\rm keV}}$ for all of our IR identified AGN and Composites. As a first step towards calculating $L_{\rm bol}$, we estimate $L_{2-10{\rm keV}}$ from $L_{\rm IR}^{\rm \scriptscriptstyle AGN}$ for all sources. We empirically determine the scaling between these luminosities to be 
\begin{align}
\label{LX}
\log \left(\frac{L_{2-10{\rm keV}}}{L_{\rm IR}^{\rm \scriptscriptstyle AGN}}\right)&= (31.698\pm3.535) \nonumber \\
&- (0.734\pm0.082)\times\log L_{\rm IR}^{\rm \scriptscriptstyle AGN}\,[{\rm erg\,s^{-1}}]
\end{align}
measured directly using {\it Chandra} observations of the GOODS-S field, which is the only field where the {\it Chandra} data is complete down to $L_{2-10{\rm keV}}= 10^{42}\,$erg\,s$^{-1}$ out to $z=2$ \citep{xue2011,hsu2014}. Note that $L_{2-10{\rm keV}}$ is the observed luminosity, as in most cases we do not have high enough counts to make a meaningful obscuration measurement. Figure \ref{xray} shows this empirically derived relationship, along with the approximate conversion factors derived in \citet{mullaney2011}, using a local sample of AGN with $L_{2-10{\rm keV}}\sim10^{43}\,$erg\,s$^{-1}$, and derived in \citet{elvis1994} from quasars with $L_{2-10{\rm keV}}>10^{45}\,$erg\,s$^{-1}$. Our conversion is in line with the literature results for the brighter AGN.

We then apply Equation \ref{LX} to all sources in the CANDELS and COSMOS fields. Next, we convert $L_{2-10{\rm keV}}$ to $L_{\rm bol}$ using Equation 2 in \citet{hopkins2007}. This Equation results in $L_{2-10{\rm keV}} / L_{\rm bol} \sim0.06-0.01$, in agreement with direct measurements in the literature \citep{vignali2003,steffan2006,vasudevan2007}. Finally, we calculate $L_{\rm Edd} [{\rm erg\,s^{-1}}]= 1.3\times 10^{38} \times M_{\rm BH} [M_\odot]$, where $M_{\rm BH} = 0.002\,M_\ast$ following the convention in \citet{marconi2003} and \citet{aird2012}.

With the techniques outlined in Section \ref{sec:colors}, we will be able to calculate $\lambda_{\rm Edd}$
for samples with $M_\ast$ or $M_{\rm BH}$ measurements. The relationship between $\lambda_{\rm Edd}$ and \fAGN\ is not linear, since $\lambda_{\rm Edd}$ depends not only on \fAGN\, but also on $L_{\rm IR}$ and $M_\ast$. Then, each \fAGN\ can have a range of $\lambda_{\rm Edd}$ depending on the host galaxy properties. We show in the top panel of Figure \ref{FOV} the distribution of 
$\lambda_{\rm Edd}$ for each galaxy category.  
In the bottom panel of Figure \ref{FOV}, we break our 10\,$\mu$m number counts into bins of $\lambda_{\rm Edd}$. Comparison with the top panel demonstrates that the $\lambda_{\rm Edd}<0.01$ curve (pink circles) is dominated by SFGs, while the $\lambda_{\rm Edd}>0.1$ curve
(yellow) has accretion rates typical of sources identified as AGN at IR and X-ray wavelengths. The majority of the counts are $\lambda_{\rm Edd}=0.01-0.1$ (purple squares), and these are objects that could be classified as AGN, SFGs, or Composites. 

The MIRI field of view is $1.2'\times1.9'$, so we also illustrate the counts in a MIRI FOV on the right axis of Figure \ref{FOV}. We expect nearly 100 objects per MIRI 
FOV down to 2\,$\mu$Jy at 10\,$\mu$m, achievable at a SNR of 10 (5) in roughly 15 minutes (3.6 minutes). Of these objects, $>$50\% may be AGN hosts where we can detect and measure the black hole accretion. Below 
$S_{10}=10\,\mu$Jy, the counts become dominated by sources with $M_\ast<10^9\,M_\odot$. Of the galaxies with $\lambda_{\rm Edd}>0.01$, 30\% have $M_\ast<10^9\,M_\odot$ and comprise a prime population for followup studies to more concretely pin down the AGN fraction in low mass galaxies at $z\sim1-2$.

The use of the $\lambda_{\rm Edd}$ parameter highlights an area where MIRI will enable great strides forward--namely, understanding how the observable properties of AGN hosts correlate to their physical properties. The broad distributions of $\lambda_{\rm Edd}$ in the top panel of Figure \ref{FOV} demonstrates the limitations of either broadly grouping sources into AGN, Composites, and SFGs based on observables, or using scaling relations to calculate physical properties, or very likely a combination of the two. But with the high resolution spectroscopy on MIRI, and the increased number of photometric filters, we will be able to classify galaxies on the relative strengths of PAH features, estimate \fAGN\, and combine with $M_\ast$ (attainable with NIRcam) to measure $\lambda_{\rm Edd}$, providing clearer insight into the relationship between galaxy dust emission and black hole accretion.

\begin{figure}
\includegraphics[width=3in]{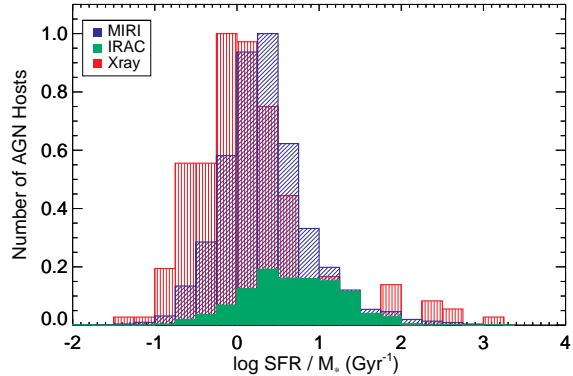}
\caption{We show the host galaxy property sSFR$=$SFR$/M_\ast$ of CANDELS galaxies identified as likely AGN according to a hard X-ray cut (red histogram; $L_{2-10{\rm keV}} \geq 10^{43}\,$erg\,s$^{-1}$) and MIRI color selection (blue histogram). We have normalized each distribution to allow easier comparison, although the MIRI distribution has 20$\times$ more sources. The MIRI selection is sensitive to host galaxies with higher sSFRs. We illustrate which of the MIRI selected AGN hosts would also be selected with the \citet{donley2012} IRAC selection criteria. Due to the sensitivity and coverage of the MIRI filters, we will be able to select larger samples of AGN hosts than was possible with IRAC selection. \label{MS}}
\end{figure}

Finally, we demonstrate the host galaxy properties of CANDELS AGN and Composites selected with different techniques at $z\sim1-2$ in Figure \ref{MS}. We calculate SFR for all galaxies by fitting templates from the \citet{kirkpatrick2015} library, based on classification as a SFG, Composite, or AGN, and then removing the AGN contribution to $L_{\rm IR}$ before converting to a SFR using Equation 3 in \citet{kennicutt1998}. We combine SFR with $M_\ast$ \citep{santini2015,stefanon2017} to measure sSFR$ = {\rm SFR} / M_\ast$, a common probe of galaxy evolution, as this ratio will be lower in galaxies that are quenching \citep[e.g][and references therein]{pandya2017}. In red, we plot the distribution of sSFR for those galaxies identified as AGN in the hard X-ray band ($L_{2-10{\rm kev}} \geq 10^{43}\,$erg\,s$^{-1}$). Then, we plot the distribution of sSFR in blue for those galaxies that will be selected as either AGN or Composites by our MIRI color diagnostics, based on our estimation of their JWST colors through template fitting. For easier comparison, we normalize both distributions to have a peak at one, although the MIRI distribution actually has 20$\times$ more galaxies than the X-ray distribution. In practice, the relative numbers of X-ray and MIRI AGN will depend on the depth of the observations and the area covered, but the sensitivity of MIRI and our ability to select Composite sources will enable larger samples than X-ray selection alone. Crucially, our cosmic noon CANDELS AGN hosts have higher sSFR than the X-ray selected CANDELS galaxies \citep{azadi2015,mullaney2015}. Combining MIRI and X-ray samples will increase our dynamic range in sSFR, allowing us to explore how black hole accretion varies with star formation and main sequence location \citep{mullaney2012,rosario2013,stanley2015}. 

Prior to JWST, the most popular way of identifying large samples of IR AGN is with IRAC color techniques \citep{lacy2004,stern2005,donley2012}. The \citet{donley2012} IRAC diagnostic 
is the most reliable, since it eliminates host galaxy contamination, but it is only sensitive to the most actively accreting AGN as it is based on a power-law selection criterion. Hence, it is likely to be significantly incomplete for Compton thick and other obscured AGN. Of the 
CANDELS sources selected by our MIRI diagnostics, we show in green the sources that are also selected as AGN by the \citet{donley2012} criteria. Clearly, due to the sensitivity and spacing of the MIRI filters, we will be able to detect $>$4 times as many AGN hosts as would be identified with IRAC alone. MIRI color selection will enable identification of statistical samples of AGN hosts in their star forming prime (as measured by sSFR), allowing astronomers to trace the star formation-AGN connection at the peak period of stellar and black hole growth in the Universe.

\section{Conclusions}
We identify SFGs, AGN, and Composites in four CANDELS fields and in the full COSMOS field using three different redshift dependent color identification techniques. We present the first 24\,$\mu$m counts of star forming+AGN Composite galaxies at $z\sim1-2$. We find that IR AGN and Composites dominate 24\,$\mu$m samples at $S_{24}>0.8\,$mJy. Any 24\,$\mu$m selected sample contains $>25\%$ of Composites.

We use a library of SFG, AGN, and Composite templates to create synthetic galaxies, and we use these synthetic galaxies to create JWST/MIRI color selection techniques for three redshift bins, $z\sim1$, $z\sim1.5$, and $z\sim2$. Our techniques can safely be applied to galaxies with $M_\ast>10^{10}\,M_\odot$. However, below this regime, metallicity may effect the strength of the PAH features, causing contamination of our Composite regime. MIRI can achieve 10$\sigma$ detections of $M_\ast<10^{10}\,M_\odot$ galaxies out to $z\sim2$ in a matter of minutes, so future JWST observations will prove crucial in separating differences in mid-IR emission due to metallicity rather than AGN in low mass galaxies.

At these redshifts, our color selection techniques cover the 6.2\,$\mu$m and 7.7\,$\mu$m PAH features and the $3-5\,\mu$m stellar minimum, which are robust tracers of star formation. We demonstrate how to correct $L_{7.7}$ for AGN contamination before converting to a SFR, a crucial step or SFRs based on 7.7\,$\mu$m PAH emission will be overestimated by $>50\%$ for AGN and $35\%$ for Composites.

Finally, we predict the Eddington ratios ($\lambda_{\rm Edd}$), a measure of black hole accretion efficiencies, that we will observe with MIRI imaging. Our MIRI color selection diagnostic can identify samples of AGN and Composite galaxies with $\lambda_{\rm Edd} > 0.01$ that are four times larger than samples of AGN selected by {\it Spitzer}/IRAC techniques. We also use our new 24\,$\mu$m number counts to predict the number counts at $10\,\mu$m in different bins of $\lambda_{\rm Edd}$. With MIRI color identification, we will be able to probe the star formation - AGN connection in dusty galaxies at cosmic noon.

\acknowledgements
A. K. thanks Sandy Faber for helpful conversations. A. K. gratefully acknowledges support
from the YCAA Prize Postdoctoral Fellowship. A. P. and A. S. acknowledge NASA ADAP13-0054 and NSF AAG grants AST- 1312418 and AST-1313206.
\clearpage
 \appendix
In this appendix, we show our redshift dependent color diagnostic to find SFGs, Composites, and AGN using {\it Spitzer} and {\it Herschel} photometry. We create a catalog of 5500 synthetic galaxies from 11 templates where we know the intrinsic AGN contribution. We resample each photometric point within the uncertainties of the template from which it was created, so that we can represent the scatter in colorspace of real galaxies, which is an improvement upon using so-called redshift tracks alone to explore where SFGs, Composites, and AGN lie in colorspace.
 
We create color diagrams in redshift bins of $z=0.75-1.25$, $z=1.25-1.75$, and $z=1.75-2.25$. In each redshift bin, we divide the color space into regions of $0.2\times0.2$\,dex and calculate the average \fAGN\ and standard deviation, $\sigma_{\rm AGN}$ of all the synthetic galaxies that lie in that region. In Figure \ref{example1}, \ref{example2}, \ref{example3} below, we show our three diagnostics: $S_{250}/S_{24}$ v. $S_8/S_{3.6}$, $S_{100}/S_{24}$ v. $S_8/S_{3.6}$, and $S_{24}/S_8$ v. $S_8/S_{3.6}$.

 \begin{figure*}[ht!]
\centering
\includegraphics[width=6in]{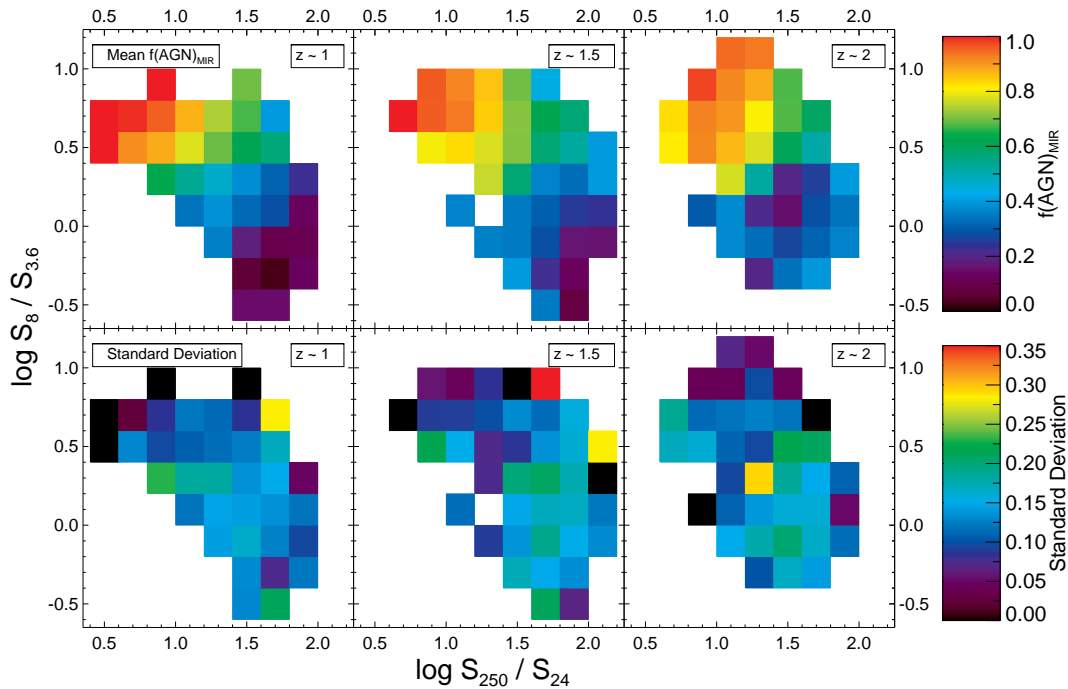}
\caption{$S_{250}/S_{24} v. S_8/S_{3.6}$ in three redshift bins. The top panels are shaded according to the average \fAGN\ measured in each bin, and the bottom panels show the standard deviation of \fAGN\ for the sources in each bin. \label{example1}}
\end{figure*}

 \begin{figure*}
\centering
\includegraphics[width=6in]{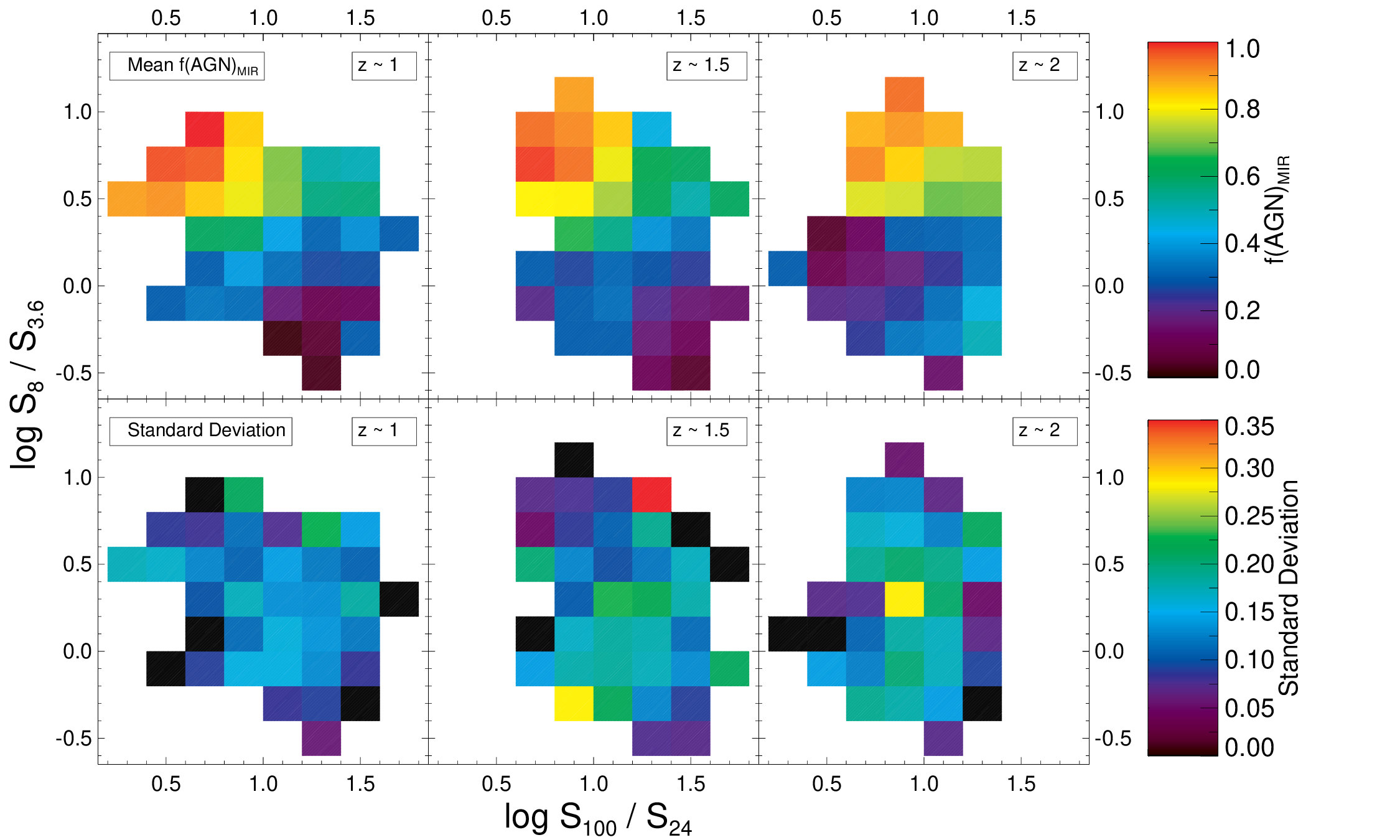}
\caption{$S_{100}/S_{24} v. S_8/S_{3.6}$ in three redshift bins. The top panels are shaded according to the average \fAGN\ measured in each bin, and the bottom panels show the standard deviation of \fAGN\ for the sources in each bin. \label{example2}}
\end{figure*}

 \begin{figure*}
\centering
\includegraphics[width=6in]{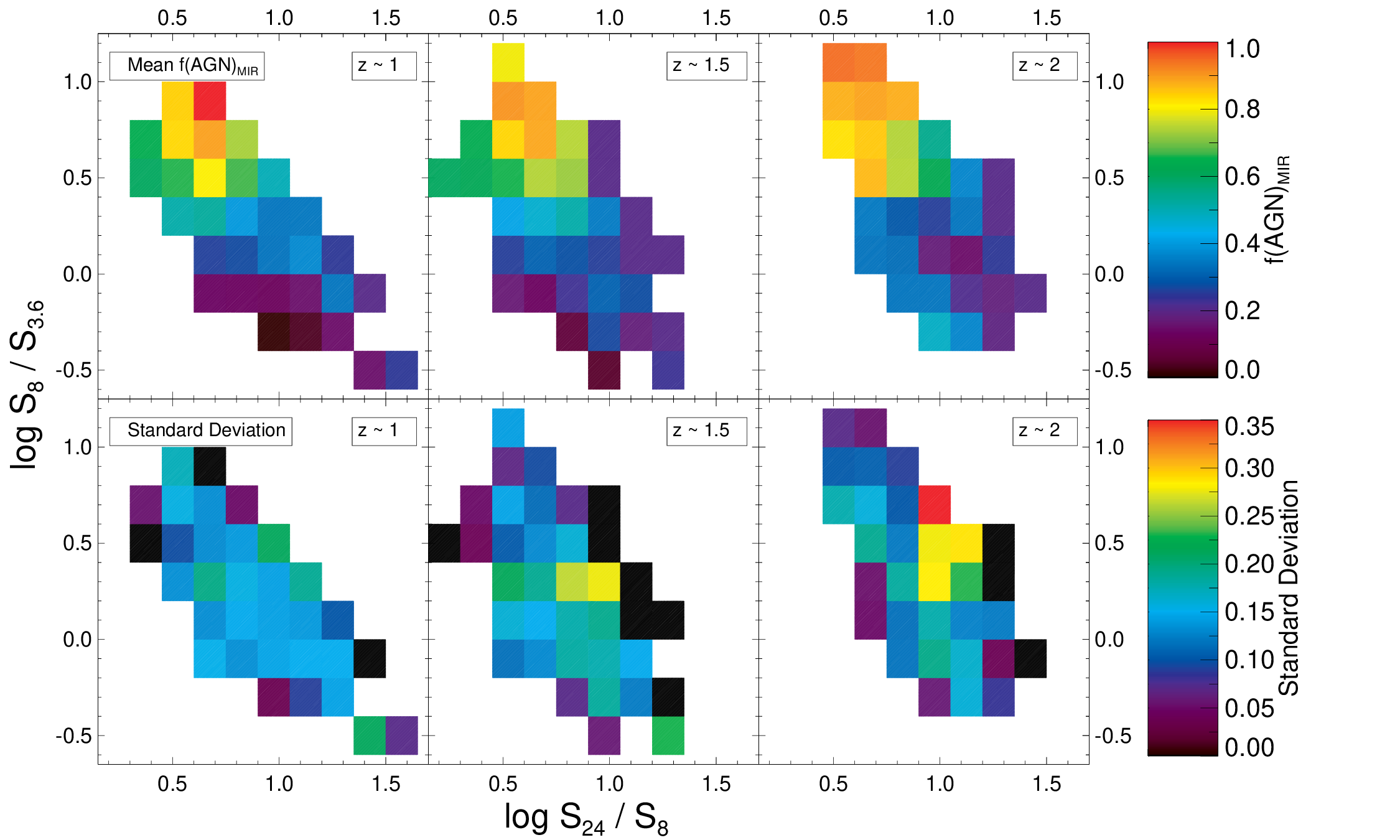}
\caption{$S_{24}/S_{8} v. S_8/S_{3.6}$ in three redshift bins. The top panels are shaded according to the average \fAGN\ measured in each bin, and the bottom panels show the standard deviation of \fAGN\ for the sources in each bin. \label{example3}}
\end{figure*}

\end{document}